\documentclass[amsmath,amssymb,nofootinbib,notitlepage,superscriptaddress,twocolumn]{revtex4-1}
\pdfoutput=1

\usepackage{aas_macros,esint,overpic}
\usepackage[bottom]{footmisc}
\usepackage[colorlinks=true]{hyperref}
\usepackage[raggedright]{subfigure}
\let\originalleft\left
\let\originalright\right
\renewcommand{\left}{\mathopen{}\mathclose\bgroup\originalleft}
\renewcommand{\right}{\aftergroup\egroup\originalright}

\begin{document}

\title{Can the EHT M87 results be used to test general relativity?}

\author{Samuel E. Gralla}
\email{sgralla@email.arizona.edu}
\affiliation{Department of Physics, University of Arizona, Tucson, Arizona 85721, USA}

\begin{abstract}
No.  All theoretical predictions for the observational appearance of an accreting supermassive black hole, as measured interferometrically by a sparse Earth-sized array at current observation frequencies, are sensitive to many untested assumptions about accretion flow and emission physics.  There is no way to distinguish a violation of general relativity (GR) from the much more likely scenario that the relevant ``gastrophysical'' assumptions simply do not hold.   Tests of GR will become possible with longer interferometric baselines (likely requiring a space mission) that reach the resolution where astrophysics-independent predictions of the theory become observable.

\end{abstract}

\maketitle

\section{Introduction}

In April 2019, the Event Horizon Telescope (EHT) collaboration released 1.3mm interferometric observations of the core of the galaxy M87 (henceforth M87*) \cite{EHT1,EHT2,EHT3,EHT4,EHT5,EHT6}.  These observations probe distance scales of order the size of the expected supermassive black hole, a remarkable technical achievement.  As we move into this exciting new era of horizon-scale electromagnetic astronomy, it is important to know what the data can---and cannot---tell us about the gravity and astrophysics of black holes.

In their initial analysis, the collaboration argued that the data are consistent with expectations for an accreting black hole, bolstering their case with a new suite of simulations of accretion flow and associated 1.3mm emission.  Using these simulations as ``ground truth'', they estimated the mass of the black hole with a reported $\sim 20\%$ uncertainty.  This result distinguished between conflicting prior mass measurements, favoring that based on stellar dynamics \cite{m87-mass-stellar} over that based on gas dynamics  \cite{m87-mass-gas}.  EHT did not claim any tests of GR or probes of accretion physics; the idea was to \textit{assume} the truth of GR and \textit{assume} truth of the emission models, and to thereby \textit{measure} the black hole mass. 

Recently, the collaboration\footnote{Not all present members of the collaboration are authors of Ref.~\cite{EHT-GR}, but the author list does state ``the EHT collaboration''.} reversed the logic \cite{EHT-GR}.  Without releasing new data or performing new analysis of existing data, they instead assumed that the stellar-dynamics mass \cite{m87-mass-stellar} is correct and argued that their original analysis \cite{EHT6} is, in fact, a test of GR.  This new approach means that EHT now has more confidence in the stellar-dynamics mass measurement than it does in the correctness of GR.  
It also means that EHT has more confidence in its accretion and emission physics assumptions than it does in the correctness of GR.  In this paper we will compare the evidence for GR with the evidence for the relevant features of the EHT emission models and conclude that GR is far better established.  This implies that no analysis requiring these emission assumptions can be used as a test of GR.  

This difficulty is not unique to the modeling and analysis choices made by EHT.  The fundamental problem is that current interferometric baselines (impressively long though they are) are not sensitive to any unique GR predictions.  The (very interesting) ``black hole shadow'' effect  \cite{falcke-melia-agol2000,bromley-melia-liu2001} is unfortunately not generic, and does not occur in the EHT models for M87*.  The ``photon ring'' due to photons that orbit the black hole \cite{luminet1979,jaroszynski-kurpiewski1997,falcke-melia-agol2000,bromley-melia-liu2001} is a generic GR prediction, but it will be very challenging to observe from Earth.  
In particular, analytical arguments \cite{gralla-holz-wald2019} supported by EHT simulations \cite{johnson-etal2020} establish that even in the most favorable of circumstances, at most $\sim10\%$ of the flux density is due to orbiting photons.\footnote{\label{sup}Ref.~\cite{johnson-etal2020} considered a single simulation from the EHT suite and reported that $\sim10\%$ of the flux comes from orbiting photons.  (This number is given at the end of section 2 and can also be inferred from the $\sim 20\%$ figure quoted in Fig.~1 together with the split by orbit number shown in Fig.~3.)  This simulation involves diffuse emission concentrated near the event horizon of a rapidly spinning black hole, which are the conditions most favorable to having a significant fraction of orbiting photons  \cite{gralla-lupsasca2020,gralla-lupsasca-marrone2020}.  Any departure from these conditions (a flatter emission profile, an emission profile ending further away from the horizon, a more equatorial emission profile, or a less rapidly spinning black hole) will decrease the orbiting photon fraction.  It would be interesting to know the orbiting photon fraction for the entire suite of EHT simulations, on which the purported GR test \cite{EHT-GR} is based.}  These photons occupy a narrow ring on the image plane (the photon ring) that is well below the effective resolution of an Earth-sized array at currently envisaged observation frequencies.  Photon ring tests of GR will almost certainly require a space mission  \cite{johnson-etal2020,gralla-lupsasca-marrone2020}.

The contention of this paper is that while the original logic of a mass estimate is reasonable, the new logic---a GR test---is not.  There is, of course, a third possibility: Assume a mass, assume GR, and test the astrophysical assumptions present in the underlying theoretical analysis.  These assumptions touch on some of the most interesting questions in relativistic astrophysics: the nature of accretion flows, the behavior of strongly magnetized plasma, and the mechanism(s) for powering and launching  relativistic jets.  It may be that the greatest benefits of continued ground-based interferometric measurements of M87* will be for a better understanding of these and other fundamental processes at work in our universe.

\section{Previous EHT claims}\label{sec:previous}

The central claims of the M87* EHT papers \cite{EHT1,EHT2,EHT3,EHT4,EHT5,EHT6} are:
\begin{enumerate}
    \item Observational Claim: The image of M87* is dominated by a ring of typical diameter $\sim 40\mu$as, whose thickness is estimated to be somewhere between $10\%$ and $50\%$ of its diameter. 
    \item Theoretical Claim: The diameter of the ring obeys an approximate scaling relation with the mass-to-distance ratio of the black hole:
\begin{align}\label{scaling}
    \theta = (40 \pm 3)\mu\textrm{as} \frac{M/(6.2 \times 10^9 M_\odot)}{D/(16.9 \textrm{ Mpc})}.
\end{align}
\end{enumerate}
Here $M$ is the black hole mass, $D$ is its distance, and $\theta$ is the angular diameter of the observed emission ring.  

\subsection{Evidence for the observational claim \\ (ring diameter)}

 Although I will not question the EHT observational claim in this paper, some readers may be interested in a summary of the evidence.  This section may be skipped without loss of continuity.

A radio interferometer measures the ``complex visibility'', which (under mild assumptions) is equal to the Fourier transform of the sky brightness.  Each pair of telescopes is sensitive to the Fourier transform at vector wavenumber equal to the telescope separation projected along the line of sight, and divided by the observation wavelength.  An $N$-telescope array thus measures $N(N-1)/2$ ``pixels'' on the Fourier plane, with additional coverage provided by the rotation of the Earth.  The fundamental problem of interferometry is to infer the properties of the image from  this limited sampling of its Fourier transform.


There are infinitely many images compatible with a given set of Fourier space pixels, so all image inferences are necessarily probabilistic.  The most straightforward method of inference is to choose a parameterized image model and fit its Fourier transform to the observed visibilities.  EHT considered a class of ``generalized crescent'' models that allow disks, rings, and crescents within a $\sim 10$ parameter freedom.  Unfortunately, the models do not fit the data until an additional $\sim 10$ nuisance parameters are included, corresponding to the addition of 2--3 Gaussian blobs on the image.  The preferred observational appearance is a narrow ring of diameter $40\mu$as (fractional width $\sim 10$--$20\%$), together with narrow bars from the nuisance Gaussians.  


EHT also considered an alternative approach, employing a class of non-parametric techniques known as imaging algorithms.  The simplest version of an imaging algorithm begins with a trial image, compares its Fourier transform to the data, and iteratively updates the image (as well as station calibration factors) based on some procedure until some specified match criterion is met.  The main drawback of this approach is that the resulting image does not come with a probabilistic interpretation.  To ascertain the reliability of the features of an image, one must rely on comparisons between the results of different algorithms run on the same data.  EHT reported that all algorithms considered revealed a ring-like feature of typical diameter $\sim40\mu$as, with fractional width varying from $\sim 30\%$ to $\sim50\%$.

Although both the parametric and non-parametric analyses have significant limitations, the limitations are somewhat orthogonal.  The fact that both classes of method robustly indicate a $\sim 40 \mu$as ring suggests that this feature is indeed present in the data.

\subsection{Evidence for the theoretical claim \\ (scaling relation)}\label{sec:evidence-for-scaling}

I have distilled the EHT theoretical claims into the scaling relation  \eqref{scaling} as follows.  I begin with the statement in Ref.~\cite{EHT6} that ``the structure and extent of the emission preferentially from outside the photon ring leads to a $\lesssim 10\%$ offset between the measured emission diameter in the model images and the size of the photon ring.''  Here by ``photon ring'' EHT means the theoretical curve on the image plane defined by the arrival of photons that have orbited the black hole arbitrarily many times before arrival, which I discuss in Sec.~\ref{sec:heuristics} below as the ``critical curve''.  This curve has a typical diameter of $37.6 \mu$as with the mass and distance used in \eqref{scaling}; I have increased this number by $7\%$ to account for the offset described by EHT, arriving at a central value of $40 \mu$as.  I have crudely estimated the claimed spread around $40\mu$as by comparing with the EHT reported posterior probability on the mass, folding in estimates of the uncertainty on the distance $D$ and the observed ring diameter $\theta$.  

The precise values present in the scaling relation \eqref{scaling} are irrelevant to the main points of this paper.  The essential point is that all EHT claims about a mass measurement or a GR test may be understood (at the order-of-magnitude level) with reference to a simple scaling relation of this form.

Let us scrutinize the evidence for the scaling relation \eqref{scaling}.  The argument given by EHT involves a bank of general relativistic magnetohydrodynamic (GRMHD) simulations using the Kerr spacetime, together with a phenomenological prescription for the emission as a function of the GRMHD variables.  I will refer to this method of generating images as GRMHD+, with the + standing for the phenomenological emission prescription.  EHT considered a large number of GRMHD+ models, varying the black hole spin, the observer inclination, the net magnetic flux on the black hole, and a parameter in their emission model.  They argued that the vast majority of these models are consistent with the data,\footnote{None of the approximately 60,000 GRMHD+ images provides a formally acceptable fit to the data (as judged by a reduced chi-squared).  This is attributed to the high variability of the flow, and an alternative ``average image scoring'' (AIS) approach was developed to determine whether the variability in a GRMHD+ simulation is statistically consistent with the observed data.  The AIS method eliminates high-magnetic-flux, retrograde flows about rapidly spinning black holes.  The use of AIS means that the scaling relation \eqref{scaling} does contain a modest amount of observational input.  However, I will continue to regard it as a theoretical result, since the majority of input comes from theoretical considerations.} but excluded some fraction based on ancillary theoretical considerations.\footnote{In the models presented in table 2 of Ref.~\cite{EHT5}, more than half were eliminated based on ancillary theoretical concerns involving jet power, X-ray luminosity, and radiative efficiency.}  The scaling relation is derived from the remaining GRMHD+ models.


Let us list the main assumptions of GRMHD+, in order from best established to least well established.  By far the best established  assumption is the Kerr metric.  GR has been tested in a variety of ways in a variety of regimes \cite{Kramer2013,Will2014,Berti2015,LIGO2016b,LIGO2017}, and the Kerr metric is an unambiguous prediction for the exterior of dark compact objects, supported by an enormous body of mathematical proof, analytical argument, and numerical evidence.

The second-best-established assumption is the fluid description of the plasma.  Given that the collisional mean free path is many orders of magnitude larger than the black hole, the validity of a fluid description is far from obvious.  
The EHT models further make the assumption of ideal MHD (infinite conductivity), a very special case in the space of plausible fluid theories of plasma.

The third-best-established assumption is the resolution of the simulations.  Global, self-consistent simulations operate at the limits of computational power, and convergence studies are challenging \cite{shiokawa-dolence-gammie-noble2012}.  Local  (shearing-box) simulations designed to study the magneto-rotational instability (MRI) are inconclusive regarding convergence, depending on the assumed magnetic environment \cite{stone-hawley-gammie-balbus1996,davis-stone-pessah2010}.  This raises some doubt as to whether the global GRMHD simulations  adequately resolve the physics that gives rise to the accretion they are designed to study.

The fourth-best-established assumption is the initial conditions for the simulations.  It is well known that the structure of the MHD solution  depends strongly on the choice of initial conditions, most particularly with regard to the formation of Poynting-flux outflows (potentially able to power a relativistic jet) \cite{beckwith-hawley-krolik2008,penna-kulkarni-akshay-narayan2013,penna-etal2010}.  The more jet-promising initial conditions are naturally used to model systems with jets (and EHT specifically excluded some of their simulations based on their lack of Poynting flux), but the strong dependence of the simulation results on the initial conditions suggests that other viable initial conditions may await discovery, especially as ideas for launching the jet continue to evolve \cite{blandford-meier-readhead2019}.

The fifth-best-established assumption is the EHT phenomenological prescription for the emission.  GRMHD follows the averaged properties of the ions, whereas it is the much-lighter electron component that is believed to give rise to the observed synchrotron emission.  Even assuming the correctness of the GRMHD results for the ion properties, the synchrotron emissivity is unknown.  EHT relied on a prescription that ties the synchrotron emissivity to the averaged properties of the ions, motivated by the idea that energy is dissipated exclusively by Landau damping.  Needless to say, this phenomenological prescription has not been tested.

The sixth-best-established assumption is the \textit{ad hoc} zeroing of emission inside the jet region.  GRMHD simulations fail inside a ``jet region'' that typically takes the shape of a paraboloid of revolution.  In this region, the density becomes so small that the GRMHD code is unable to continue the evolution.  To avoid this difficulty, simulations impose an arbitrary floor on the density, which forces the dynamics to be approximately force-free.  While perhaps reliable for determining the evolution of the large-scale electromagnetic field, this approach is incapable of predicting the emission from the small-scale particle acceleration and pair-creation that is believed to occur \cite{blandford-znajek1977}.  The EHT collaboration chose to make the emission zero in this region.


\section{Heuristic Interpretations}\label{sec:heuristics}

The EHT GRMHD+ models predict a relatively tight scaling \eqref{scaling} between black hole mass and observed ring diameter.  EHT suggests that this agreement in observational appearance among their GRMHD+ models is a result of gravitational lensing, invoking the heuristics of a ``shadow'' and a ``photon ring''.  In this section I will review these and other heuristics used to interpret black hole images and argue that the agreement in observational appearance cannot plausibly be attributed to a shadow or photon ring.  Instead, the similar observational appearance of the EHT GRMHD+ models likely means that all have very similar emission profiles, at least when projected along the line of sight.


\subsection{Backlit shadow}\label{sec:backlit-shadow}

A black hole illuminated from behind by an isotropic source of much larger angular radius will cast a dark shadow of size somewhat larger than the critical curve, inside of which will be a sequence of thin rings converging to the critical curve \cite{gralla-holz-wald2019}.  If the illumination is instead from the entire celestial sphere (including behind the observer), then the shadow shrinks down to the critical curve.  These scenarios are irrelevant for understanding the appearance of matter near black holes, and indeed were not considered by EHT.

\subsection{Doppler shadow}\label{sec:doppler-shadow}

A black hole surrounded by optically thin, radially infalling matter will appear as a dark hole whose outline is the critical curve (Fig.~1(a) of Ref.~\cite{falcke-melia-agol2000}).\footnote{Ref.~\cite{falcke-melia-agol2000} also considers matter orbiting on Keplerian shells [Fig.~1(d)].  This model produces a more gentle decrease not clearly associated with the critical curve.}  This effect is caused by extreme Doppler deboosting \cite{narayan-johnson-gammie2019} due to the assumption of radial infall.  Since rays arriving within the critical curve by definition have no turning point, all photons in that area were emitted opposite to the radially infalling flow and suffer strong Doppler deboosting.


This effect was named the black hole ``shadow''.  The terminology is somewhat unfortunate, since the phenomenon has nothing to do with backlighting, instead being caused by special-relativistic Doppler deboosting \cite{narayan-johnson-gammie2019}.  A more appropriate name would perhaps be ``Doppler deficit''.  Meeting halfway between these terms, I will  use the name ``Doppler shadow'' for the dimming inside the critical curve due to radial infall~\cite{falcke-melia-agol2000,narayan-johnson-gammie2019}.



\subsection{Discrete photon rings}\label{sec:discrete-photon-rings}

The existence of unstable photon orbits gives rise to multiple images of sources near black holes \cite{darwin1959}.  In principle there are infinitely many such images, indexed by the total number of orbits and the direction of the orbit (clockwise or counter-clockwise).  The images accumulate near the critical curve and are increasingly demagnified.   When the source is in the shape of a disk, the high-order images are thin rings (``photon rings'') converging to the critical curve \cite{luminet1979,hollywood-melia1997,beckwith-done2005}.  For optically thin disks extending to near the horizon (as presumed to occur in M87*), the photon rings comprise a distinctive multi-peak structure \cite{gralla-holz-wald2019} that is seen in (at least one of) the EHT GRMHD+ simulations when ray-traced at sufficiently high resolution \cite{johnson-etal2020}.  The photon rings contain at most $\sim 10\%$ of the total flux in such a simulation (see footnote \ref{sup}), consistent with theoretical expectations \cite{gralla-holz-wald2019}.

\subsection{Continuous photon ring}

When the source is diffuse (with no sharp features like point sources or disks) as well as optically thin, the heuristic of multiple images becomes less useful.  Instead, we may consider the observed intensity to be proportional to the optical path of a ray traced backward through the source.  The optical path diverges at the critical curve, leading to a smooth brightness enhancement \cite{jaroszynski-kurpiewski1997,falcke-melia-agol2000}.  However, the enhancement is only logarithmic \cite{gralla-holz-wald2019,johnson-etal2020,gralla-lupsasca2020}.



\subsection{Just add one}\label{sec:just-add-one}

Although full ray-tracing in the Kerr metric is a computationally intensive process, some of its features can be understood in simple terms.  Consider, for example, the question of which point on the image corresponds to which region of the source.  If our source is a disk (thin or thick) centered on the spin-equator of the black hole, we may ask for the relationship between the equatorial emission radius $r$ (using Boyer-Lindquist coordinates) and the arrival position on the image.  In the face-on limit (suitable as a first approximation for M87*), we may use radial impact parameter $b$ to represent radial distance on the image.  The answer for the direct photons (which do not orbit) is shockingly simple:
\begin{align}
    b \approx r+M.
\end{align}
That is, you ``just add one''.  This was noticed numerically in Ref.~\cite{gralla-lupsasca2020} and derived as an analytic approximation in Ref.~\cite{gates-hadar-lupsasca2020}.  We may interpret the formula as stating that the primary image of the disk is essentially pristine, with no significant distortion due to lensing, and no influence of the black hole spin.\footnote{This lack of distortion persists in the modestly inclined case---see, for example, the second column of Fig.~6 of Ref.~\cite{gralla-lupsasca2020}.}

\subsection{Interpretation of GRMHD+ images}\label{sec:EHT-models}

Which heuristics are most useful for understanding the EHT GRMHD+ images?  The backlit shadow is not relevant (and was not considered by EHT).  The Doppler shadow is also not relevant, as the dark region can be significantly displaced from the critical curve.\footnote{See, for example, the middle panel of Fig.~1 of Ref.~\cite{EHT5}, noting the contrast between the thin ring (the photon ring, tracing the critical curve) and the dark region in the center.}  The narrow photon ring feature is present in all images, but it does not contain enough flux to significantly affect the observed ring diameter (see footnote \ref{sup}).  Based on the striking qualitative agreement with purely equatorial toy models \cite{gralla-holz-wald2019,gralla-lupsasca-marrone2020} (for example, compare Fig.~1 of Ref.~\cite{gralla-holz-wald2019} to Fig.~3 of Ref.~\cite{johnson-etal2020}), it appears that the most useful heuristics for interpreting the  GRMHD+ images are ``just add one'' (Sec.~\ref{sec:just-add-one}) for the majority of the flux, together with discrete photon rings (secondary images) appearing near the critical curve (Sec.~\ref{sec:discrete-photon-rings}).

In particular, gravitational lensing plays little role in determining the observed ring diameter at EHT resolution.  Instead, the majority of the flux is in the relatively undistorted ``primary image'' of the source,  composed of the integrated emissivity along direct (non-orbiting) rays through the disk.   That is, the structure of the observed ring is determined by the projected structure of the source disk.

\section{Alternative Scenarios}\label{sec:alternative-scaling}


I have argued that the scaling relation \eqref{scaling} is not due to gravitational lensing, but rather a direct consequence of the assumed source structure.  It is instructive to consider alternative source structures under which the scaling relation would be significantly altered.  In fact, precisely such a scenario was given by the antecedents of EHT in 2010 \cite{doeleman2012}.  This paper presented the first evidence of a $\sim 40 \mu$\textrm{as} 1.3mm structure in M87*.  To interpret their results, the authors argued that the size of the emission region can be associated with the innermost stable circular orbit (ISCO) of the black hole.  Translated into an effective scaling relation, the theoretical claims of \cite{doeleman2012} entail an effective allowed range for $\theta$,\footnote{For these estimates I use the range $1M$--$9M$ for the Boyer-Lindquist radius of the ISCO, together with the fact that the relevant lensing boils down to ``just adding one'' (Sec.~\ref{sec:just-add-one}).}
\begin{align}
    \theta_{\rm min} & = 14\mu\textrm{as} \frac{M/(6.2 \times 10^9 M_\odot)}{D/(16.9 \textrm{ Mpc})} \\
    \theta_{\rm max} & = 72\mu\textrm{as} \frac{M/(6.2 \times 10^9 M_\odot)}{D/(16.9 \textrm{ Mpc})}.
\end{align}
As explained by the authors of \cite{doeleman2012}, this ``ISCO scenario'' is natural because the mass density drops at the ISCO (even if the disk is geometrically thick, with significant stresses).  The ISCO scenario predicts a wide range $\theta_{\rm min} < \theta < \theta_{\rm max}$ for the emission size, depending on black hole spin.  By constrast, the EHT GRMHD+ scenario \eqref{scaling} predicts a very narrow range of $\theta$, independent of black hole spin.

Another alternative to the GRMHD+ scenario is the idea that the observed emission comes not from an accretion flow but from a near-horizon magnetosphere.  This idea was discussed recently by Blandford, Meier, and Readhead \cite{blandford-meier-readhead2019}.  After reviewing some issues with the standard interpretation of the EHT observations, the authors state that ``a simpler hypothesis...is that EHT is observing dynamically insignificant relativistic plasma orbiting with the angular velocity of an ordered magnetic field, confined by a much larger ejection disk.''  Although the emission from such a scenario has not yet been calculated, it would presumably come from very near the horizon and therefore produce a scaling relation with a smaller prefactor than Eq.~\eqref{scaling}.  

Finally, even within the broad paradigm of an accretion flow extending down to the horizon \cite{melia1992,narayan-yi1994}, some studies have suggested that non-thermal emission from near the horizon (not included in GRMHD+) may play an important---or even dominant---role in the millimeter-wave observational appearance of M87* \cite{broderick-loeb2009,levinson-rieger2011,moscibrodzka2011,broderick-tchekhovskoy2015,hirotani-pu2016}.




\section{Summary and Conclusions}\label{sec:summary}

The scaling relation \eqref{scaling} claimed by EHT is derived from a bank of GRMHD+ simulations (Sec.~\ref{sec:evidence-for-scaling}).  The scaling relation is very nearly identical to that which would arise by identifying the observed emission ring with the critical curve (also sometimes called the ``shadow'' or ``photon ring'').  This is an accident of the particular assumptions in the EHT models of the source.  There is no shadow effect in the EHT simulations used to derive the scaling relation (Sec.~\ref{sec:EHT-models}), and the photon ring contributes minimally to the structure of these (or any other) images on the effective resolution of Earth-sized baselines (Sec.~\ref{sec:discrete-photon-rings}).  The scaling relation \eqref{scaling} is not a universal prediction of GR, but a specific prediction of a specific class of source models.

On general grounds, this means that no analysis using the scaling relation can be considered a test of GR.  The gastrophysical assumptions underlying the scaling relation (Sec.~\ref{sec:evidence-for-scaling}) are \textit{far} less well-established than general relativity.  Alternative scaling relations are conceivable and indeed have been considered in the past by some of the authors of \cite{EHT-GR} (Sec.~\ref{sec:alternative-scaling}).  Given the overwhelming evidence for GR and the much more limited evidence for astrophysical assumptions used in deriving any given scaling relation, null-hypothesis testing can only probe astrophysics, not gravity.

Beyond the null-hypothesis idea, the recent paper \cite{EHT-GR} attempts to place constraints on alternative metrics.  The assumption is that, when these alternative metrics are considered, the new scaling relation will be determined by identifying the observed ring with the critical curve of that new metric.  This assumption is valid in GR only in special cases, and no justification for it holding in alternative theories is given in Ref.~\cite{EHT-GR}.  
Even if an MHD+ analysis were performed with an alternative metric to determine an alternative scaling relation, there would be no way to distinguish a failure of the gravity theory from a failure of the astrophysical assumptions.

In conclusion, the diameter of the observed ring in M87*, as determined by the present EHT observations and analysis, cannot be used to test GR.

\section*{Acknowledgements}

This work was supported in part by NSF grant PHY-1752809 to the University of Arizona.

\bibliography{noGRtest}

\providecommand{\href}[2]{#2}\begingroup\raggedright\begin{thebibliography}{10}

\bibitem{EHT1}
{Event Horizon Telescope Collaboration}, K.~{Akiyama}, A.~{Alberdi}, {\em
  et~al.}, ``{First M87 Event Horizon Telescope Results. I. The Shadow of the
  Supermassive Black Hole},''
  \href{http://dx.doi.org/10.3847/2041-8213/ab0ec7}{{\em \apjl} {\bfseries 875}
  (Apr., 2019) L1}.

\bibitem{EHT2}
{Event Horizon Telescope Collaboration}, K.~{Akiyama}, A.~{Alberdi}, {\em
  et~al.}, ``{First M87 Event Horizon Telescope Results. II. Array and
  Instrumentation},'' \href{http://dx.doi.org/10.3847/2041-8213/ab0c96}{{\em
  \apjl} {\bfseries 875} (Apr., 2019) L2}.

\bibitem{EHT3}
{Event Horizon Telescope Collaboration}, K.~{Akiyama}, A.~{Alberdi}, {\em
  et~al.}, ``{First M87 Event Horizon Telescope Results. III. Data Processing
  and Calibration},'' \href{http://dx.doi.org/10.3847/2041-8213/ab0c57}{{\em
  \apjl} {\bfseries 875} (Apr., 2019) L3}.

\bibitem{EHT4}
{Event Horizon Telescope Collaboration}, K.~{Akiyama}, A.~{Alberdi}, {\em
  et~al.}, ``{First M87 Event Horizon Telescope Results. IV. Imaging the
  Central Supermassive Black Hole},''
  \href{http://dx.doi.org/10.3847/2041-8213/ab0e85}{{\em \apjl} {\bfseries 875}
  (Apr., 2019) L4}.

\bibitem{EHT5}
{Event Horizon Telescope Collaboration}, K.~{Akiyama}, A.~{Alberdi}, {\em
  et~al.}, ``{First M87 Event Horizon Telescope Results. V. Physical Origin of
  the Asymmetric Ring},''
  \href{http://dx.doi.org/10.3847/2041-8213/ab0f43}{{\em \apjl} {\bfseries 875}
  (Apr., 2019) L5}.

\bibitem{EHT6}
{Event Horizon Telescope Collaboration}, K.~{Akiyama}, A.~{Alberdi}, {\em
  et~al.}, ``{First M87 Event Horizon Telescope Results. VI. The Shadow and
  Mass of the Central Black Hole},''
  \href{http://dx.doi.org/10.3847/2041-8213/ab1141}{{\em \apjl} {\bfseries 875}
  (Apr., 2019) L6}.

\bibitem{m87-mass-stellar}
K.~{Gebhardt}, J.~{Adams}, D.~{Richstone}, {\em et~al.}, ``{The Black Hole Mass
  in M87 from Gemini/NIFS Adaptive Optics Observations},''
  \href{http://dx.doi.org/10.1088/0004-637X/729/2/119}{{\em \apj} {\bfseries
  729} (Mar., 2011) 119}, \href{http://arxiv.org/abs/1101.1954}{{\ttfamily
  arXiv:1101.1954}}.

\bibitem{m87-mass-gas}
J.~L. {Walsh}, A.~J. {Barth}, L.~C. {Ho}, and M.~{Sarzi}, ``{The M87 Black Hole
  Mass from Gas-dynamical Models of Space Telescope Imaging Spectrograph
  Observations},'' \href{http://dx.doi.org/10.1088/0004-637X/770/2/86}{{\em
  \apj} {\bfseries 770} (June, 2013) 86},
  \href{http://arxiv.org/abs/1304.7273}{{\ttfamily arXiv:1304.7273}}.

\bibitem{EHT-GR}
D.~{Psaltis}, L.~{Medeiros}, P.~{Christian}, {\em et~al.}, ``{Gravitational
  Test Beyond the First Post-Newtonian Order with the Shadow of the M87 Black
  Hole},'' {\em arXiv e-prints} (Oct., 2020) arXiv:2010.01055,
  \href{http://arxiv.org/abs/2010.01055}{{\ttfamily arXiv:2010.01055 [gr-qc]}}.

\bibitem{falcke-melia-agol2000}
H.~{Falcke}, F.~{Melia}, and E.~{Agol}, ``{Viewing the Shadow of the Black Hole
  at the Galactic Center},'' \href{http://dx.doi.org/10.1086/312423}{{\em
  \apjl} {\bfseries 528} (Jan., 2000) L13--L16},
  \href{http://arxiv.org/abs/astro-ph/9912263}{{\ttfamily astro-ph/9912263}}.

\bibitem{bromley-melia-liu2001}
B.~C. {Bromley}, F.~{Melia}, and S.~{Liu}, ``{Polarimetric Imaging of the
  Massive Black Hole at the Galactic Center},''
  \href{http://dx.doi.org/10.1086/322862}{{\em \apjl} {\bfseries 555} no.~2,
  (July, 2001) L83--L86},
  \href{http://arxiv.org/abs/astro-ph/0106180}{{\ttfamily
  arXiv:astro-ph/0106180 [astro-ph]}}.

\bibitem{luminet1979}
J.-P. {Luminet}, ``{Image of a spherical black hole with thin accretion
  disk},'' {\em \aap} {\bfseries 75} (May, 1979) 228--235.

\bibitem{jaroszynski-kurpiewski1997}
M.~{Jaroszynski} and A.~{Kurpiewski}, ``{Optics near Kerr black holes: spectra
  of advection dominated accretion flows.},'' {\em \aap} {\bfseries 326} (Oct.,
  1997) 419--426, \href{http://arxiv.org/abs/astro-ph/9705044}{{\ttfamily
  astro-ph/9705044}}.

\bibitem{gralla-holz-wald2019}
S.~E. {Gralla}, D.~E. {Holz}, and R.~M. {Wald}, ``{Black hole shadows, photon
  rings, and lensing rings},''
  \href{http://dx.doi.org/10.1103/PhysRevD.100.024018}{{\em \prd} {\bfseries
  100} no.~2, (July, 2019) 024018},
  \href{http://arxiv.org/abs/1906.00873}{{\ttfamily arXiv:1906.00873
  [astro-ph.HE]}}.

\bibitem{johnson-etal2020}
M.~D. Johnson, A.~Lupsasca, A.~Strominger, {\em et~al.}, ``Universal
  interferometric signatures of a black hole{\textquoteright}s photon ring,''
  \href{http://dx.doi.org/10.1126/sciadv.aaz1310}{{\em Science Advances}
  {\bfseries 6} no.~12, (2020) }.

\bibitem{gralla-lupsasca2020}
S.~E. {Gralla} and A.~{Lupsasca}, ``{Lensing by Kerr black holes},''
  \href{http://dx.doi.org/10.1103/PhysRevD.101.044031}{{\em \prd} {\bfseries
  101} no.~4, (Feb., 2020) 044031},
  \href{http://arxiv.org/abs/1910.12873}{{\ttfamily arXiv:1910.12873 [gr-qc]}}.

\bibitem{gralla-lupsasca-marrone2020}
S.~E. {Gralla}, A.~{Lupsasca}, and D.~P. {Marrone}, ``{The Shape of the Black
  Hole Photon Ring: A Precise Test of Strong-Field General Relativity},'' {\em
  arXiv e-prints} (Aug., 2020) arXiv:2008.03879,
  \href{http://arxiv.org/abs/2008.03879}{{\ttfamily arXiv:2008.03879 [gr-qc]}}.

\bibitem{Kramer2013}
M.~{Kramer}, \href{http://dx.doi.org/10.1017/S174392131202306X}{``{Probing
  gravitation with pulsars},''} in {\em Neutron Stars and Pulsars: Challenges
  and Opportunities after 80 years}, J.~{van Leeuwen}, ed., vol.~291 of {\em
  IAU Symposium}, pp.~19--26.
\newblock Mar., 2013.
\newblock \href{http://arxiv.org/abs/1211.2457}{{\ttfamily arXiv:1211.2457
  [astro-ph.HE]}}.

\bibitem{Will2014}
C.~M. {Will}, ``{The Confrontation between General Relativity and
  Experiment},'' \href{http://dx.doi.org/10.12942/lrr-2014-4}{{\em Living
  Reviews in Relativity} {\bfseries 17} no.~1, (Dec., 2014) 4},
  \href{http://arxiv.org/abs/1403.7377}{{\ttfamily arXiv:1403.7377 [gr-qc]}}.

\bibitem{Berti2015}
E.~{Berti}, E.~{Barausse}, V.~{Cardoso}, {\em et~al.}, ``{Testing general
  relativity with present and future astrophysical observations},''
  \href{http://dx.doi.org/10.1088/0264-9381/32/24/243001}{{\em Classical and
  Quantum Gravity} {\bfseries 32} no.~24, (Dec., 2015) 243001},
  \href{http://arxiv.org/abs/1501.07274}{{\ttfamily arXiv:1501.07274 [gr-qc]}}.

\bibitem{LIGO2016b}
B.~P. {Abbott}, R.~{Abbott}, T.~D. {Abbott}, {\em et~al.}, ``{Tests of General
  Relativity with GW150914},''
  \href{http://dx.doi.org/10.1103/PhysRevLett.116.221101}{{\em \prl} {\bfseries
  116} no.~22, (June, 2016) 221101},
  \href{http://arxiv.org/abs/1602.03841}{{\ttfamily arXiv:1602.03841 [gr-qc]}}.

\bibitem{LIGO2017}
B.~P. {Abbott}, R.~{Abbott}, T.~D. {Abbott}, {\em et~al.}, ``{Gravitational
  Waves and Gamma-Rays from a Binary Neutron Star Merger: GW170817 and GRB
  170817A},'' \href{http://dx.doi.org/10.3847/2041-8213/aa920c}{{\em \apjl}
  {\bfseries 848} no.~2, (Oct., 2017) L13},
  \href{http://arxiv.org/abs/1710.05834}{{\ttfamily arXiv:1710.05834
  [astro-ph.HE]}}.

\bibitem{shiokawa-dolence-gammie-noble2012}
H.~{Shiokawa}, J.~C. {Dolence}, C.~F. {Gammie}, and S.~C. {Noble}, ``{Global
  General Relativistic Magnetohydrodynamic Simulations of Black Hole Accretion
  Flows: A Convergence Study},''
  \href{http://dx.doi.org/10.1088/0004-637X/744/2/187}{{\em \apj} {\bfseries
  744} no.~2, (Jan., 2012) 187},
  \href{http://arxiv.org/abs/1111.0396}{{\ttfamily arXiv:1111.0396
  [astro-ph.HE]}}.

\bibitem{stone-hawley-gammie-balbus1996}
J.~M. {Stone}, J.~F. {Hawley}, C.~F. {Gammie}, and S.~A. {Balbus},
  ``{Three-dimensional Magnetohydrodynamical Simulations of Vertically
  Stratified Accretion Disks},'' \href{http://dx.doi.org/10.1086/177280}{{\em
  \apj} {\bfseries 463} (June, 1996) 656}.

\bibitem{davis-stone-pessah2010}
S.~W. {Davis}, J.~M. {Stone}, and M.~E. {Pessah}, ``{Sustained
  Magnetorotational Turbulence in Local Simulations of Stratified Disks with
  Zero Net Magnetic Flux},''
  \href{http://dx.doi.org/10.1088/0004-637X/713/1/52}{{\em \apj} {\bfseries
  713} no.~1, (Apr., 2010) 52--65},
  \href{http://arxiv.org/abs/0909.1570}{{\ttfamily arXiv:0909.1570
  [astro-ph.HE]}}.

\bibitem{beckwith-hawley-krolik2008}
K.~{Beckwith}, J.~F. {Hawley}, and J.~H. {Krolik}, ``{The Influence of Magnetic
  Field Geometry on the Evolution of Black Hole Accretion Flows: Similar Disks,
  Drastically Different Jets},'' \href{http://dx.doi.org/10.1086/533492}{{\em
  \apj} {\bfseries 678} no.~2, (May, 2008) 1180--1199},
  \href{http://arxiv.org/abs/0709.3833}{{\ttfamily arXiv:0709.3833
  [astro-ph]}}.

\bibitem{penna-kulkarni-akshay-narayan2013}
R.~F. {Penna}, A.~{Kulkarni}, and R.~{Narayan}, ``{A new equilibrium torus
  solution and GRMHD initial conditions},''
  \href{http://dx.doi.org/10.1051/0004-6361/201219666}{{\em \aap} {\bfseries
  559} (Nov., 2013) A116}, \href{http://arxiv.org/abs/1309.3680}{{\ttfamily
  arXiv:1309.3680 [astro-ph.HE]}}.

\bibitem{penna-etal2010}
R.~F. {Penna}, J.~C. {McKinney}, R.~{Narayan}, {\em et~al.}, ``{Simulations of
  magnetized discs around black holes: effects of black hole spin, disc
  thickness and magnetic field geometry},''
  \href{http://dx.doi.org/10.1111/j.1365-2966.2010.17170.x}{{\em \mnras}
  {\bfseries 408} no.~2, (Oct., 2010) 752--782},
  \href{http://arxiv.org/abs/1003.0966}{{\ttfamily arXiv:1003.0966
  [astro-ph.HE]}}.

\bibitem{blandford-meier-readhead2019}
R.~{Blandford}, D.~{Meier}, and A.~{Readhead}, ``{Relativistic Jets from Active
  Galactic Nuclei},''
  \href{http://dx.doi.org/10.1146/annurev-astro-081817-051948}{{\em \araa}
  {\bfseries 57} (Aug., 2019) 467--509},
  \href{http://arxiv.org/abs/1812.06025}{{\ttfamily arXiv:1812.06025
  [astro-ph.HE]}}.

\bibitem{blandford-znajek1977}
R.~D. {Blandford} and R.~L. {Znajek}, ``{Electromagnetic extraction of energy
  from Kerr black holes},'' {\em \mnras} {\bfseries 179} (May, 1977) 433--456.

\bibitem{narayan-johnson-gammie2019}
R.~{Narayan}, M.~D. {Johnson}, and C.~F. {Gammie}, ``{The Shadow of a
  Spherically Accreting Black Hole},'' {\em arXiv e-prints} (Oct, 2019)
  arXiv:1910.02957, \href{http://arxiv.org/abs/1910.02957}{{\ttfamily
  arXiv:1910.02957 [astro-ph.HE]}}.

\bibitem{darwin1959}
C.~{Darwin}, ``{The Gravity Field of a Particle},''
  \href{http://dx.doi.org/10.1098/rspa.1959.0015}{{\em Proceedings of the Royal
  Society of London Series A} {\bfseries 249} (Jan., 1959) 180--194}.

\bibitem{hollywood-melia1997}
J.~M. {Hollywood} and F.~{Melia}, ``{General Relativistic Effects on the
  Infrared Spectrum of Thin Accretion Disks in Active Galactic Nuclei:
  Application to Sagittarius A *},''
  \href{http://dx.doi.org/10.1086/313036}{{\em \apjs} {\bfseries 112} no.~2,
  (Oct., 1997) 423--455}.

\bibitem{beckwith-done2005}
K.~{Beckwith} and C.~{Done}, ``{Extreme gravitational lensing near rotating
  black holes},''
  \href{http://dx.doi.org/10.1111/j.1365-2966.2005.08980.x}{{\em \mnras}
  {\bfseries 359} (June, 2005) 1217--1228},
  \href{http://arxiv.org/abs/astro-ph/0411339}{{\ttfamily astro-ph/0411339}}.

\bibitem{gates-hadar-lupsasca2020}
D.~E.~A. {Gates}, S.~{Hadar}, and A.~{Lupsasca}, ``{Maximum Observable
  Blueshift from Circular Equatorial Kerr Orbiters},'' {\em arXiv e-prints}
  (Sept., 2020) arXiv:2009.03310,
  \href{http://arxiv.org/abs/2009.03310}{{\ttfamily arXiv:2009.03310 [gr-qc]}}.

\bibitem{doeleman2012}
S.~S. {Doeleman}, V.~L. {Fish}, D.~E. {Schenck}, {\em et~al.}, ``{Jet-Launching
  Structure Resolved Near the Supermassive Black Hole in M87},''
  \href{http://dx.doi.org/10.1126/science.1224768}{{\em Science} {\bfseries
  338} no.~6105, (Oct., 2012) 355},
  \href{http://arxiv.org/abs/1210.6132}{{\ttfamily arXiv:1210.6132
  [astro-ph.HE]}}.

\bibitem{melia1992}
F.~{Melia}, ``{An Accreting Black Hole Model for Sagittarius A *},''
  \href{http://dx.doi.org/10.1086/186297}{{\em \apjl} {\bfseries 387} (Mar.,
  1992) L25}.

\bibitem{narayan-yi1994}
R.~{Narayan} and I.~{Yi}, ``{Advection-dominated Accretion: A Self-similar
  Solution},'' \href{http://dx.doi.org/10.1086/187381}{{\em \apjl} {\bfseries
  428} (June, 1994) L13},
  \href{http://arxiv.org/abs/astro-ph/9403052}{{\ttfamily
  arXiv:astro-ph/9403052 [astro-ph]}}.

\bibitem{broderick-loeb2009}
A.~E. {Broderick} and A.~{Loeb}, ``{Imaging the Black Hole Silhouette of M87:
  Implications for Jet Formation and Black Hole Spin},''
  \href{http://dx.doi.org/10.1088/0004-637X/697/2/1164}{{\em \apj} {\bfseries
  697} no.~2, (June, 2009) 1164--1179},
  \href{http://arxiv.org/abs/0812.0366}{{\ttfamily arXiv:0812.0366
  [astro-ph]}}.

\bibitem{levinson-rieger2011}
A.~{Levinson} and F.~{Rieger}, ``{Variable TeV Emission as a Manifestation of
  Jet Formation in M87?},''
  \href{http://dx.doi.org/10.1088/0004-637X/730/2/123}{{\em \apj} {\bfseries
  730} no.~2, (Apr., 2011) 123},
  \href{http://arxiv.org/abs/1011.5319}{{\ttfamily arXiv:1011.5319
  [astro-ph.HE]}}.

\bibitem{moscibrodzka2011}
M.~{Mo{\'s}cibrodzka}, C.~F. {Gammie}, J.~C. {Dolence}, and H.~{Shiokawa},
  ``{Pair Production in Low-luminosity Galactic Nuclei},''
  \href{http://dx.doi.org/10.1088/0004-637X/735/1/9}{{\em \apj} {\bfseries 735}
  no.~1, (July, 2011) 9}, \href{http://arxiv.org/abs/1104.2042}{{\ttfamily
  arXiv:1104.2042 [astro-ph.HE]}}.

\bibitem{broderick-tchekhovskoy2015}
A.~E. {Broderick} and A.~{Tchekhovskoy}, ``{Horizon-scale Lepton Acceleration
  in Jets: Explaining the Compact Radio Emission in M87},''
  \href{http://dx.doi.org/10.1088/0004-637X/809/1/97}{{\em \apj} {\bfseries
  809} no.~1, (Aug., 2015) 97},
  \href{http://arxiv.org/abs/1506.04754}{{\ttfamily arXiv:1506.04754
  [astro-ph.HE]}}.

\bibitem{hirotani-pu2016}
K.~{Hirotani} and H.-Y. {Pu}, ``{Energetic Gamma Radiation from Rapidly
  Rotating Black Holes},''
  \href{http://dx.doi.org/10.3847/0004-637X/818/1/50}{{\em \apj} {\bfseries
  818} no.~1, (Feb., 2016) 50},
  \href{http://arxiv.org/abs/1512.05026}{{\ttfamily arXiv:1512.05026
  [astro-ph.HE]}}.

\end{thebibliography}\endgroup
\bibliographystyle{utphys}

\end{document}